\documentclass[pra,twocolumn,showpacs,preprintnumbers,amsmath,amssymb]{revtex4}
\usepackage{graphicx,graphics,color,times,slashed}
\begin{document}
\title{Geometric phases and quantum phase transitions}
\author{Shi-Liang Zhu}
\email{slzhu@scnu.edu.cn} \affiliation{Institute for Condensed
Matter Physics, School of Physics and Telecommunication
Engineering, South China Normal University, Guangzhou, China}

\begin{abstract}
Quantum phase transition is one of the main interests in the field
of condensed matter physics, while geometric phase is a
fundamental concept and has attracted considerable interest in the
field of quantum mechanics. However, no relevant relation was
recognized before recent work. In this paper, we present a review
of the connection recently established between these two
interesting fields: investigations in the geometric phase of the
many-body systems have revealed so-called "criticality of
geometric phase", in which geometric phase associated with the
many-body ground state exhibits universality, or scaling behavior
in the vicinity of the critical point. In addition, we address the
recent advances on the connection of some other geometric
quantities and quantum phase transitions. The closed relation
recently recognized between quantum phase transitions and some of
geometric quantities may open attractive avenues and fruitful
dialog between different scientific communities.
\end{abstract}


\maketitle

\section{Introduction}

Quantum phase transition (QPT), which is closely associated with
the fundamental changes that can occur in the macroscopic nature
of matter at zero temperature due to small variations in a given
external parameter, is certainly one of the major interests in
condensed matter physics. Actually, the past decade has seen a
substantial rejuvenation of interest in the study of quantum phase
transition, driven by experiments on the cupric superconductors,
the heavy fermion materials, insulator-superfluid transition in
ultrocold atoms, organic conductors and related
compounds\cite{Sachdev,Wen}. Quantum phase transitions are
characterized by the dramatic changes in the ground state
properties of a system driven by quantum fluctuations.
Traditionally phases and phase transitions are described by the
Ginzburg-Landau symmetry-breaking theory based on order parameters
and long range correlation. Recently, substantially effort has
been devoted to the analysis of quantum phase transitions from
other intriguing perspectives, such as topological
order\cite{Wen}, quantum entanglement\cite{Osterloh,Gu}, geometric
phases\cite{Carollo,Zhu2006} and some other geometric
quantities\cite{Zanardi0,Zanardi1,Zhou,Venuti,Quan}.

It is well-known that geometric ideas have played an important
role in physics. For example, Minkiwski's geometric reformulation
of special relativity by means of a space-time geometry was very
useful in the construction of general relativity by Einstein. In
this paper we will address another example: the study of quantum
phase transition from the perspective of  geometric phase (GP)
factors. Actually,the phase factor of a wave function is the
source of all interference phenomena and one of most fundamental
concepts in quantum physics. The first considerable progress in
this field is achieved  by Aharonov and Bohm in
1959\cite{Aharonov59}. They proposed that the loop integral of the
electromagnetic potentials gives an observed nonintegrable phase
factor in electron interference experiments. By using the
non-Abelian phase factor, Yang reformulated the concept of gauge
fields in an integral formalism in 1974\cite{Yang74}, and then Wu
and Yang showed that the gauge phase factor gives an intrinsic and
complete description of electromagnetism. It neither
underdescribes nor overdescribes it\cite{Wu_Yang}. The recent
considerable interests in this field are motivated by a pioneer
work by Berry in 1984\cite{Berry}, where he discovered that a
geometric phase, in addition to the usual dynamical phase, is
accumulated on the wave function of a quantum system, provided
that the Hamiltonian is cyclic and adiabatic. It was Simon who
first recognized the deep geometric meaning underlying Berry's
phase. He observed that geometric phase is what mathematicians
would call a $U(1)$ holonomy in the parameter space, and the
natural mathematical context for holonomy is the theory of fiber
bundles\cite{Simon}. A further important generalization of Berry's
concept was introduced by Aharonov and Anandan\cite{Aharonov87},
provided that the evolution of the state is cyclic. Besides,
Samuel and Bhandari introduced a more general geometric phase in
the nonadiabatic noncyclic evolution of the system\cite{Samuel}.
Now the applications of Berry phases and its generalizations
\cite{Berry,Simon,Aharonov87,Samuel,Sjoqvist,Zhu2000} can be found
in many physical fields, such as optics, magnetic resonance,
molecular and atomic physics, condensed matter physics and quantum
computation, {\sl
etc}.\cite{Shapere,Li98,Bohm,Thouless,Morpurgo,Zanardi}.

Very recently, investigations in the geometric phase of the
many-body systems have revealed so-called "criticality of
geometric phase"\cite{Carollo,Zhu2006}, in which geometric phase
associated with the ground state exhibits universality, or scaling
behavior, around the critical point\cite{Zhu2006}. The closed
relation between quantum phase transitions and geometric phases
may be understood from an intuitive view: quantum phase
transitions occur for a parameter region where the energy levels
of the ground state and the excited state cross or have an avoided
crossing, while geometric phase, as a measure of the curvature of
Hilbert space, can reflect the energy structures and then can
capture certain essential features of quantum phase
transitions\cite{Zhu2006}.

A typical example to show the significant connection between
geometric phase and quantum phase transition is one-dimensional XY
spin chain\cite{Carollo,Zhu2006}. Since the XY spin chain model is
exactly solvable and still presents a rich structure, it has
become a benchmark to test many new concepts. The XY spin chain
model and the geometric phase that corresponds to the quantum
phase transition have been analyzed in detail in
Ref.\cite{Carollo,Zhu2006}. The XY model is parameterized by
$\gamma$ and $\lambda$ (see the definitions below
Eq.(\ref{Hamiltonian})). Two distinct critical regions appear in
parameter space: the segment $(\gamma,\lambda)=(0,(0,1))$ for the
XX chain and the critical line $\lambda_c=1$ for the whole family
of the XY model\cite{Sachdev,Lieb}. It has been shown that
geometric phase can be used to characterize the above two critical
regions\cite{Carollo,Zhu2006,Hamma}. As for the first critical
region, a noncontractible geometric phase
itself\cite{Zhu2006,Hamma} or its difference between the ground
state and the first excited state\cite{Carollo} exists in the XX
chain if and only if the closed evolution path circulates a region
of criticality. There are much more physics in the second critical
region since second order quantum phase transition occur there.
The geometric phase of the ground state has been shown to have
scaling behavior near the critical point of the XY model. In
particular, it has been found that the geometric phase is
non-analytical and its derivative with respect to the field
strength $\lambda$ diverges logarithmically near the critical line
described by $\lambda_c=1$. Together with a logarithmic divergence
of the derivative as a function of system size, the critical
exponents are derived based on the scaling ansatz in the case of
logarithmic divergence\cite{Barber}. Furthermore, universality in
the critical properties of geometric phase for a family of XY
models is verified.  These results show that the key ingredients
of quantum criticality are present in the ground-state geometric
phase and therefore are indicators of criticality of geometric
phase\cite{Zhu2006}.

\begin{figure}[tbph]
\centering
\includegraphics[height=5cm]{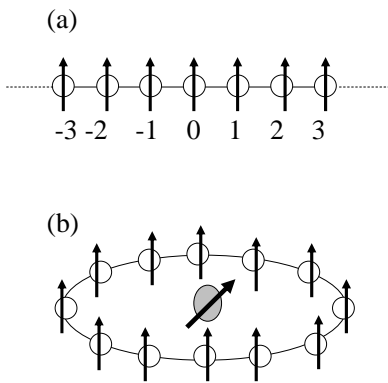}
\caption{Schematic diagrams of the physical patterns reviewed in
the paper. (a) Pattern I:  N spins in one-dimensional chain is the
whole system. The geometric phase of the whole N-spin system has
close relation with quantum phase transitions of the whole system.
(b) Pattern II: N spins are arranged in a circle and a test qubit
in the center possesses homogeneous coupling with all N-spin in
the ring. The geometric phase of the test qubit may be used to
locate the criticality of quantum phase transition exhibits in the
N-spin system. Depending on the couplings between the spins,
N-spin chain (ring) in (a) and (b) can be classified as the XY
model, the Dicke model and the Lipkin-Meshkoc-Glick model. All
these three models exhibit quantum phase transitions which
features can be captured by the geometric phases or some other
geometric quantities.} \label{Fig1}
\end{figure}

Motivated by these results in the XY model\cite{Carollo,Zhu2006},
the criticality of geometric phase for other many-body models are
investigated\cite{Plastina,Chen,Cui,Yi,Yuan,Cozzini}. Roughly
speaking, there are two patterns (see Fig.1) in literature to
investigate the criticality of geometric phase in the many-body
systems: (i) Pattern I is order to investigate the relation
between geometric phase of the whole many-body system and the
system's quantum phase transition. As illustrate in Fig.1 (a), the
geometric phase of the whole N-spin system is calculated and its
scaling features in the vicinity of critical points are
discussed\cite{Carollo,Zhu2006,Hamma,Plastina,Chen,Cui}. (b)
Pattern II is concerned with the geometric phase of a test qubit
as shown in Fig. 1(b).  N spins are arranged in a circle and a
test qubit in the center possesses homogeneous coupling with all
N-spin in the ring\cite{Quan,Yi,Yuan}. The geometric phase of the
test qubit may be used to locate the criticality of quantum phase
transition exhibiting in the N-spin system\cite{Yi,Yuan}.
Depending on the couplings between the spins, N-spin chain (ring)
in (a) and (b) can be classified as the XY model, the Dicke model
and the Lipkin-Meshkoc-Glick model. All these three models exhibit
quantum phase transitions, whose features can be captured by the
geometric phases in both patterns I and II.

Furthermore, the study of QPTs by using other geometric
quantities, such as quantum overlap (quantum
fidelity)\cite{Zanardi0}, the Riemannian tensor\cite{Zanardi1}
etc., has been put forward and fruitful results have been reported
in literature. In particular, GP is a imagine part of quantum
geometric tensor and quantum fidelity is a real part, therefore a
unified theory of study QPTs from the perspective of quantum
geometric tensor has been developed\cite{Venuti}.

In this paper we will review some aspects of the theoretical
understanding that has emerged over the past several years towards
understanding the close relation between GPs and QPTs. In section
2, we present the connection between Berry curvature and QPs.
Section 3 describes the detailed relation between QPT and GP in
the patter I. Section 4 discusses the results in the patter II.
Finally, Section 5 presents some discussion and perspective in the
topic reviewed in this paper, in particular, we address the recent
advances in the connection of some other geometric quantities and
QPTs.

\section{Berry curvature and quantum phase transitions}

Let us first address the close relation between quantum phase
transitions and geometric phases from an intuitive view. Consider
a generic many-body system described by the Hamiltonian $H(\eta)$
with $\eta$ a dimensionless coupling constant. For any reasonable
$\eta$, all observable properties of the ground state of $H$ will
vary smoothly as $\eta$ is varied. However, there may be special
points denoted as $\eta_c$, where there is a non-analyticity in
some properties of the ground state at zero temperature, $\eta_c$
is identified as the position of a quantum phase transition.
Non-analytical behavior generally occur at level crossings or
avoided level crossings\cite{Sachdev}. Surprisingly, the geometric
phase is able to capture such kinds of level structures and is
therefore expected to signal the presence of quantum phase
transitions. To address this relation in greater detail, we review
geometric phases in a generic many-body system where the
Hamiltonian can be changed by varying the parameters ${\bf R}$ on
which it depends. The state $|\psi (t)\rangle$ of the system
evolves according to Schrodinger equation
\begin{equation}
\label{Schrodinger} i\hbar
\partial_t |\psi (t)\rangle=H ({\bf R} (t))|\psi(t)\rangle.
\end{equation}
At any instant, the natural basis consists of the eigenstates
$|n({\bf R})\rangle$ of $H({\bf R})$ for ${\bf R}={\bf R}(t)$,
that satisfy $H ({\bf R})|n ({\bf R})\rangle=E_n ({\bf R})|n ({\bf
R})\rangle$ with energy $E_n ({\bf R})$ $(n=1,2,3\cdots)$. Berry
showed that the GP for a specific eigenstate, such as the ground
state ($|g\rangle=|1\rangle$) of a many-body system we concern
here, adiabatically undergoing a closed path in parameter space
denoted by $C$, is given by\cite{Berry}
\begin{equation}\beta_g
(C)=-\int\int_C V_g ({\bf R})\cdot d{\bf S},
\end{equation} where
$d{\bf S}$ denotes area element in ${\bf R}$ space and $V_g ({\bf
R})$ is the Berry curvature given by
\begin{equation}
\label{Curvarure} V_g ({\bf R})=Im\sum_{n \not= g}\frac{\langle
g|\nabla_{\bf R} H|n\rangle \langle n|\nabla_{\bf R}
H|g\rangle}{(E_n-E_g)^2}.
\end{equation}
The energy denominators in Eq.(\ref{Curvarure}) show that the
Berry curvature usually diverges at the point of parameter space
where energy levels are cross and may have maximum values at
avoided level crossings. Thus level crossings or avoided level
crossings (seem Fig. 2), the two specific level structures related
to quantum phase transitions, are reflected in the geometry of the
Hilbert space of the system and can be captured by the Berry
curvature of the ground state. However, although the Berry
curvature is gauge invariant and is therefore an observable
quantity, no feasible experimental setup has been proposed to
directly observe it. On the other hand, the area integral of Berry
curvature, i.e., the geometric phase may be measured by the
interference experiments. Therefore, rather than the Berry
curvature, hereafter we will focus on the relation between
geometric phase and quantum phase transition, and therefore the
proposed relation between them may be experimentally tested.

\begin{figure}[tbph]
\centering \label{Fig2}
\includegraphics[height=4cm,width=9cm]{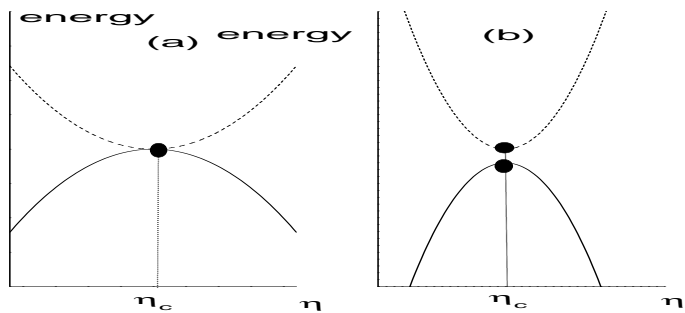}
\caption{Schematic representation of the energy level for the
many-body systems. The energy levels of the ground state and the
excited state cross in (a) and have an avoided crossing in (b). On
the one hand, quantum phase transition occurs at level crossings
or avoided level crossings, which represents by the parameter
$\eta_c$; on the other hand, the Berry curvature usually diverges
or may have maximum values at the point of parameter $\eta_c$. }
\end{figure}

\section{Pattern I: QPT and GP of the many-body systems}

In this section we review the closed relation between QPTs and GPs
for the Pattern I, as shown in Fig.1 (a), where the N-spin chain
can be classified as the XY model, the Dicke model and the
Lipkin-Meshkoc-Glick model.

\subsection{The XY spin chain}

Our first example is one-dimensional XY spin chain investigated in
detail in Ref.\cite{Zhu2006} . The XY model concerns N spin-1/2
particles (qubits) with nearest neighbor interactions and an
external magnetic field. The Hamiltonian of the XY spin chain has
the following form
\begin{equation}
\label{Hamiltonian}  H=-\sum_{j=-M}^M \left
(\frac{1+\gamma}{2}\sigma_j^x\sigma_{j+1}^x+\frac{1-\gamma}{2}\sigma_j^y\sigma_{j+1}^y+\lambda\sigma_j^z
\right), \end{equation} where $\sigma^\mu_j\ (\mu=x,y,x)$ are the
Pauli matrices for the $j$th spin, $\gamma$ represents the
anisotropy in the $x-y$ plane and $\lambda$ is the intensity of
the magnetic field applied in the $z$ direction. We assume
periodic boundary conditions for simplicity and choose $N\
(=2M+1)$ odd to avoid the subtleties connected with the boundary
terms. Nevertheless, the differences with other boundary
conditions and the even $N$ case are the order to O(1/N) and then
negligible in the thermodynamic limit where quantum phase
transitions occur\cite{Lieb,Osterloh}. This XY model encompasses
two other well-known spin models: it turns into transverse Ising
chain for $\gamma=1$ and the XX (isotropic XY) chain in a
transverse field for $\gamma=0$.

In order to derive the geometric phase of ground state in this
system, we introduce a new family of Hamiltonians that can be
described by applying a rotation of $\phi$ around the $z$
direction to each spin \cite{Carollo}, {\sl i.e.},

\begin{equation}H_\phi=U_\phi^\dagger H U_\phi,\ \ U_\phi=\prod_{j=-M}^M
\exp(-i\phi\sigma_j^z/2).\end{equation} The critical behavior is
independent of $\phi$ as the spectrum $\Lambda_k$ (see below) of
the system is $\phi$ independent. This class of models can be
diagonalized by means of the Jordan-Wigner transformation that
maps spins to one-dimensional spinless fermions with creation and
annihilation operators $a_j$ and $a_j^\dagger$ via the relations,
$a_j=(\prod_{l<j} \sigma_l^z)\sigma_j^\dagger$
\cite{Lieb,Sachdev}. Due to the (quasi) translational symmetry of
the system we may introduce Fourier transforms of the fermionic
operator described by $d_k=\frac{1}{\sqrt{N}}\sum_j a_j
\exp(-i2\pi jk/N)$ with $k=-M,\cdots,M$. The Hamiltonian $H_\phi$
can be diagonalized by transforming the fermion operators in
momentum space and then using the standard Bogoliubov
transformation. In this way, we obtain the following diagonalized
form of the Hamiltonian,
\begin{equation}H=\sum_k \Lambda_k
(c_k^\dagger c_k-1),\end{equation} where the energy of one
particle excitation is given by
\begin{equation}\Lambda_k=\sqrt{(\lambda-\cos(2\pi
k/N))^2+\gamma^2\sin^2(2\pi k/N)}\end{equation} and $c_k=d_k
\cos\frac{\theta_k}{2}-id_{-k}^\dagger e^{2i\phi}
\sin\frac{\theta_k}{2}$ with the angle $\theta_k$ defined by
$\cos\theta_k=(\cos\frac{2\pi k}{N}-\lambda)/\Lambda_k$.

The ground state $|g\rangle$ of $H_\phi$ is the vacuum of the
fermionic modes described by $c_k |g\rangle=0$. Substituting the
operator $c_k$ into this equation, one obtains the ground state as
\begin{equation} |g\rangle=\prod_{k=1}^M \left (
\cos\frac{\theta_k}{2}|0\rangle_k |0\rangle_{-k} -i e^{2i\phi}
\sin\frac{\theta_k}{2}|1\rangle_k|1\rangle_{-k} \right),
\end{equation}
 where $|0\rangle_k$ and $|1\rangle_k$ are the
vacuum and single excitation of the $k$th mode, respectively. The
ground state is a tensor product of states, each lying in the
two-dimensional Hilbert space spanned by
$|0\rangle_k|0\rangle_{-k}$ and $|1\rangle_k|1\rangle_{-k}$. The
geometric phase of the ground state, accumulated by varying the
angle $\phi$ from $0$ to $\pi$ (Because the Hamiltonian $H_\phi$
has bilinear form, $H_\phi$ is $\pi$ periodic in $\phi$ ), is
described by
\begin{equation}\beta_g=-\frac{i}{M}\int_0^\pi \langle g|\partial_\phi
|g\rangle d\phi.\end{equation}  The direct calculation
shows\cite{Carollo}

\begin{equation}
\label{Phase} \beta_g=\frac{\pi}{M}\sum_{k=1}^M (1- \cos\theta_k).
\end{equation}
The term $\beta_k\equiv \pi (1-\cos\theta_k)$ is a geometric phase
for the $k$th mode, and represents the area in the parameter space
(which is the Bloch sphere) enclosed by the loop determined by
$(\theta_k,\phi)$. To study the quantum criticality, we are
interested in the thermodynamic limit when the spin lattice number
$N\ \to \infty$. In this case the summation
$\frac{1}{M}\sum_{k=1}^M$ can be replaced by the integral
$\frac{1}{\pi}\int_0^\pi d\varphi$ with $\varphi=\frac{2\pi
k}{N}$; and then the geometric phase in the thermodynamic limit is
given by
\begin{equation}
\label{Limit} \beta_g=\int_0^\pi  (1-\cos\theta_\varphi)d\varphi,
\end{equation}
where $\cos\theta_\varphi=(\cos\varphi-\lambda)/\Lambda_\varphi$
with the energy spectrum
$\Lambda_{\varphi}=\sqrt{(\lambda-\cos\varphi)^2+\gamma^2\sin^2\varphi}$.

As for quantum criticality in the XY model, there are two regions
of criticality, defined by the existence of gapless excitations in
the parameter space $(\gamma,\lambda)$: (i) the XX region of
criticality described by the segment $(\gamma,\lambda)=(0,(0,1))$;
(ii) the critical line $\lambda_c=1$ for the whole family of the
XY model. For the second critical region, we need to distinguish
two universality classes depending on the anisotropy $\gamma$. The
critical features are characterized in term of a critical exponent
$\nu$ defined by $\xi \sim |\lambda-\lambda_c|^{-\nu}$ with $\xi$
representing the correlation length. For any value of $\gamma $,
quantum criticality occurs at a critical magnetic field
$\lambda_c=1$. For the interval $0<\gamma\le 1$ the models belong
to the Ising universality class characterized by the critical
exponent $\nu=1$, while for $\gamma=0$ the model belongs to the XX
universality class with $\nu=1/2$ \cite{Lieb,Sachdev}. The close
relation between geometric phase and quantum criticality for the
first region has been addressed in
Refs.\cite{Carollo,Zhu2006,Hamma}, here we mainly review the
results for the second region, which is clearly more interesting
in the sense that the second order quantum phase transitions occur
there.

\begin{figure}[tbph]
\centering
\includegraphics[height=5cm,width=8cm]{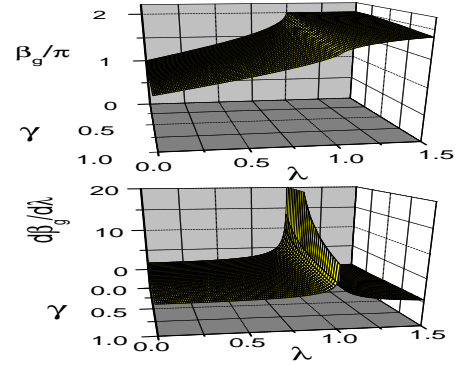}
\caption{(color online). (a) Geometric phase $\beta_g$ of the
ground state (b) and its derivative $d\beta_g/d\lambda$ as a
function of the Hamiltonian parameters $\lambda$ and $\gamma$. The
lattice size $N=10001$. There are clear anomalies for the
derivative of geometric phase along the critical line
$\lambda_c=1$. } \label{Fig3}
\end{figure}

To demonstrate the relation between geometric phase and quantum
phase transitions, we plot geometric phase $\beta_g$ and its
derivative $d\beta_g/d\lambda$ with respect to the field strength
$\lambda$ and $\gamma$ in Fig.3. A significate feature is notable:
the nonanalytical property of the geometric phase along the whole
critical line $\lambda_c=1$ in the XY spin model is clearly shown
by anomalies for the derivative of geometric phase along the same
line.

\begin{figure}[tbph]
\centering \vspace{1.5cm}
\includegraphics[height=3.5cm,width=8cm]{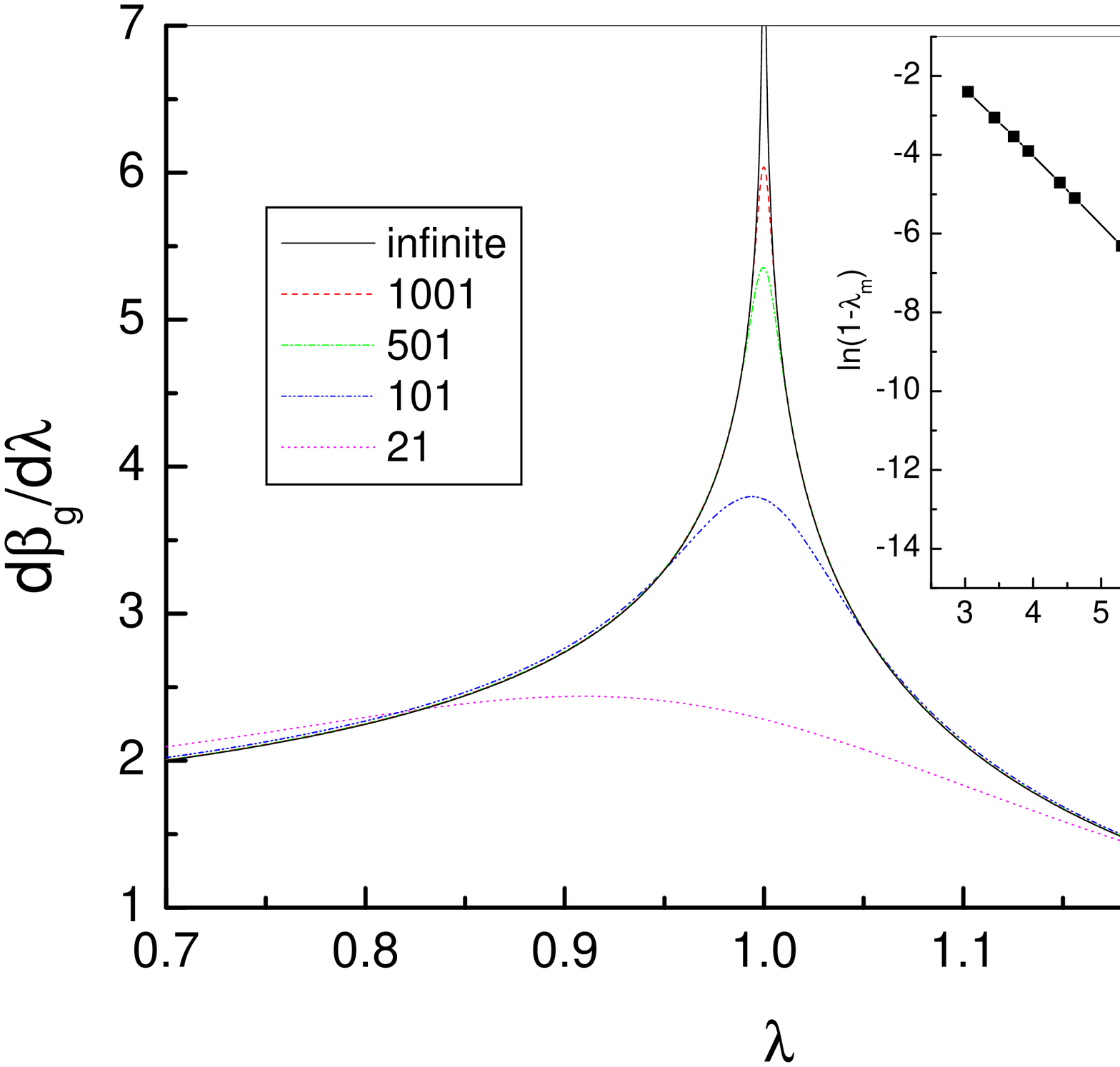}
\vspace{-1.0cm}
\caption{(color online). The derivatives $d\beta_g/d\lambda$ for
the Ising model ($\gamma=1$) as a function of the Hamiltonian
parameter $\lambda$. The curves correspond to different lattice
sizes $N=21,101,501,1001,\infty$. With increasing the system
sizes, the maximum becomes more pronounced. The inset shows that
the position of the maximum changes and tends as $N^{-1.803}$
towards the critical point $\lambda_c=1$. } \label{Fig4}
\end{figure}

To further understand the relation between geometric phase and
quantum criticality, we study the scaling behavior of geometric
phases by the finite size scaling approach\cite{Barber}. We first
look at the Ising model. The derivatives $d\beta_g/d\lambda$ for
$\gamma=1$ and different lattice sizes are plotted in
Fig.\ref{Fig4}. There is no real divergence for finite $N$, but
the curves exhibit marked anomalies and the height of which
increases with lattice size. The position $\lambda_m$ of the peak
can be regarded as a pseudo-critical point \cite{Barber} which
changes and tends as $N^{-1.803}$ towards the critical point and
clearly approaches $\lambda_c$ as $N \to \infty$. In addition, as
shown in Ref \cite{Zhu2006}, the value of $d\beta_g/d\lambda$ at
the point $\lambda_m$ diverges logarithmically with increasing
lattice size as:
\begin{equation}
\label{Scaling1} \frac{d\beta_g}{d\lambda}|_{\lambda_m} \approx
\kappa_1 \ln N +\mbox{const.},
\end{equation}
with $\kappa_1=0.3121$. On the other hand, the singular behavior
of $d\beta_g/d\lambda$ for the infinite Ising chain can be
analyzed in the vicinity of the quantum criticality, and we find
the asymptotic behavior as
\begin{equation} \label{Scaling2}
\frac{d\beta_g}{d\lambda}\approx \kappa_2
 \ln |\lambda-\lambda_c|+\mbox{ const.}, \end{equation}
with $\kappa_2=-0.3123$. According to the scaling ansatz in the
case of logarithmic divergence \cite{Barber}, the ratio
$|\kappa_2/\kappa_1|$ gives the exponent $\nu$ that governs the
divergence of the correlation length. Therefore, $\nu \sim 1$ is
obtained in our numerical calculation for the Ising chain, in
agreement with the well-known solution of the Ising model
\cite{Lieb}.

A cornerstone of QPTs is a universality principle in which the
critical behavior depends only on the dimension of the system and
the symmetry of the order parameter. The XY model for the interval
$\gamma \in (0,1]$ belong to the same universality class with
critical exponent $\nu=1$. To verify the universality principle in
this model, the scaling behavior for different values of the
parameter $\gamma$ has been numerically calculated in Ref.
\cite{Zhu2006}. The results there shown that the asymptotic
behaviors are still described by Eqs. (\ref{Scaling1}) and
(\ref{Scaling2}) with $\kappa_1$ and $\kappa_2$ being
$\gamma$-dependent constants, and the same critical exponent
$\nu=1$ can be obtained for any $\gamma \in (0,1]$.

Comparing with the $\gamma\not= 0$ case, the nature of the
divergence of $d\beta_g/d\lambda$ at the critical point
$(\gamma=0,\lambda=1)$ belongs to a different universality class,
and the scaling behavior of geometric phase can be directly
extracted from the explicit expression of the geometric phase in
the thermodynamic limit. The geometric phase under the
thermodynamic limit can be obtained explicitly from
Eq.(\ref{Limit}) for $\gamma=0$ as

\begin{eqnarray}
\beta_g  =\left\{
\begin{array}
{ll}%
2\pi, & (\lambda \leq 1)\\
2\pi-2 \arccos(\lambda), & (\lambda > 1)%
\end{array}
\right.
\end{eqnarray}
 However, it appears from
Eq.(\ref{Phase}) that the geometric phase $\beta_g$ is always
trivial for strictly $\gamma=0$ and every finite lattice size $M$,
since $\theta_k=0$ or $\pi$ for every $k$. The difference between
the finite and infinite lattice sizes  can be understood from the
two limits $N\to \infty$ and $\gamma \to 0$. Assume
$\gamma=\epsilon$ with $\epsilon$ an arbitrary small but still
finite value, then we can still find a solution $\varphi_0$ (it
implies $N\to \infty$) for $\cos\varphi_0-\lambda=0$ but $\Lambda
_{\varphi_0}=\epsilon\sqrt{1-\lambda^2}\neq 0$ for $\lambda\neq
1$. Then a $\pi$ geometric phase appears for such $\varphi_0$
since $\theta_{\varphi_0}=\pi/2$. Since
$d\beta_g/d\lambda=\sqrt{2}(1-\lambda)^{-1/2}$ $(\lambda \to
1^{-})$, we can infer the known result that the critical exponent
$\nu=1/2$ for the XX model.

Furthermore, we can confirm the known equivalent $z\nu=1$ between
$\nu$ and the dynamical exponent $z$  from the calculations of
geometric phases. The dynamical behavior is determined by the
expansion of the energy spectrum, i.e.,  $\Lambda_{\varphi
\rightarrow 0} \sim \varphi^z [1+(\varphi\xi)^{-z}]$. Then $z=1$
for $\gamma\in(0,1]$ and $z=2$ for $\gamma=0$ are found by the
expansion of $\Lambda_\varphi$ in the case $\varphi\rightarrow 0$.
So we have $z\nu=1$, which is indeed the case for the XY
criticality\cite{Sachdev}.

Therefore, the above results clearly show that all the key
ingredients of the quantum criticality are present in the
geometric phases of the ground state in the XY spin model.

\subsection{The Dicke model}

Our second example is the Dicke model \cite{Dicke} studied in
Ref.\cite{Plastina,Chen}. It consists of $N$ two-level (qubit)
systems coupled to a single Bosonic mode. The Hamiltonian is given
by ($\hbar=1$)

\begin{equation}
H=\omega a^{+}a+\Delta
J_x+\frac{\lambda}{\sqrt{N}}(a^\dagger+a)J_z,
\end{equation}
where $a$, $a^{+}$ are the annihilation and creation operators of
the Bosonic mode, respectively;
$J_{x,z}=\sum_{j=1}^{N}\sigma_{x,z}^j$ with $\sigma_{x,z}^j$ being
the Pauli matrices for the qubit $j$ are collective angular
momentum operators for all qubits; $\lambda$ denotes the coupling
strength between the atom and field; The parameters $\Delta$ and
$\omega$ represent the transition frequency of the atom and
Bosonic mode frequency, respectively. The prefactor $1/\sqrt{N}$
is inserted to have a finite free energy per atom in the
thermodynamical limit $N\rightarrow\infty$. This Hamiltonian is
canonically equivalent to the Dicke Hamiltonian  by a $\pi/2$
rotation around the $y$ axis.

As illustrated in Refs.\cite{Hepp1,Emary}, exact solutions may be
obtained in the thermodynamic limit by employing a
Holstein-Primakoff transformation of the angular momentum algebra.
In the thermodynamical limit, the Dicke Hamiltonian undergoes a
second quantum phase transition at the critical point
$\lambda_c=\sqrt{\omega\Delta/2}.$ When $\lambda<\lambda_c$, the
system is in its normal phase in which the ground state is highly
unexcited, while $\lambda>\lambda_c$, the system is in its
superradiant phase in which both the bosonic field occupation and
the spin magnetization acquire macroscopic values.

Similarly to the XY spin model, in order to investigate the
geometric phase one changes the original Hamiltonian by the
unitary transformation $U_\phi=\exp(-i\phi J_x/2)$ where $\phi$ is
a slowly varying parameter, and then the transformed Hamiltonian
is given by
\begin{equation}
\label{Dicke} H_\phi=U^\dagger_\phi H
U_\phi=\frac{\omega}{2}[p^2+q^2+\mathbf{B}\cdot \mathbf{J}],
\end{equation}
where the Hamiltonian of the free bosonic field is expressed in
terms of canonical variables $q=(a^\dagger+a)/\sqrt{2}$ and
$p=i(a^\dagger-a)/\sqrt{2}$ that obey the standard quantization
condition $[q,p]=i$.
$\mathbf{B}=(D,\frac{Lq}{\sqrt{N}\sin\phi},\frac{Lq}{\sqrt{N}\cos\phi})$
with dimensionless parameters $D=2\Delta/\omega$ and
$L=2\sqrt{2}\lambda/\omega$ is an effective magnetic field felt by
the qubits.

In the adiabatic limit, the geometric phase associated with the
ground state of the system can be obtained by the Born-Oppenheimer
approximation \cite{Plastina,Liberti}. In this case, the total
wave function of the ground state of the system can be
approximated by
\begin{equation}
\label{Ground_state} |\psi_{tot}\rangle=\int dq \varphi
(q)|q\rangle \otimes |\chi (q,\phi)\rangle.
\end{equation}
Here the state $|\chi (q,\varphi\rangle$ is the state of the
adiabatic equation of the qubit ("fast") part for each fixed value
of the slow variable $q$, i.e.,
\begin{equation}
\mathbf{B} \cdot \mathbf{J} |\chi (q,\varphi)\rangle=E (q) |\chi
(q,\varphi)\rangle
\end{equation}
with $E(q)$ the eigenenergy. It can be proven that the state
$|\chi (q,\varphi)\rangle$ can be expressed as a direct product of
$N$ qubits as $|\chi (q,\varphi)\rangle=\otimes_{j=1}^N |\chi
(q,\varphi)\rangle_j,$ and the state of each qubit can be written
as $$|\chi
(q,\varphi)\rangle_j=\sin\frac{\alpha}{2}|\uparrow\rangle_j-\cos\frac{\alpha}{2}
e^{-i\eta} |\downarrow\rangle_j$$ with $\cos\alpha=Lq
\cos\phi/(\sqrt{N} E(q))$ and $\tan\eta=Lq \sin\phi/(\sqrt{N} D).$
On the other hand, the ground state wave function for the
oscillator $\varphi (q)$ is governed by one-dimensional
time-independent Schrodinger equation
$$
H_{ad} |\varphi (q)\rangle=\frac{\omega}{2} \left( \frac{d^2}{d
q^2}+q^2-N E(q) \right)=\varepsilon_0 |\varphi (q)\rangle,
$$
where $\varepsilon_0$ is the lowerest eigenvalues of the adiabatic
Hamiltonian $H_{ad}$.

Once the total wave function of the ground state is derived, the
geometric phase $\beta_g$ of the ground state may be derived by
the standard method as $\beta_g=i\oint
\langle\psi_{tot}|d/d\phi|\psi_{tot}\rangle d\varphi$, and the
final result is given by
\begin{equation}
\beta_g=N\pi \left(1+\frac{\langle J_x\rangle}{N}
\right).\end{equation} In the thermodynamic limit, one can show
that
\begin{equation}
\frac{\beta_g}{N} \left|_{N\to\infty} \right.=\left\{
\begin{array}{ll}\ \ \ \ 0, & (\alpha \leq 1 ) \\ \pi
(1-\frac{1}{\alpha}),\ \ \ \ & (\alpha > 1 ).\end{array} \right.
\end{equation}

\begin{figure}[tbph]
\centering
\includegraphics[height=4cm]{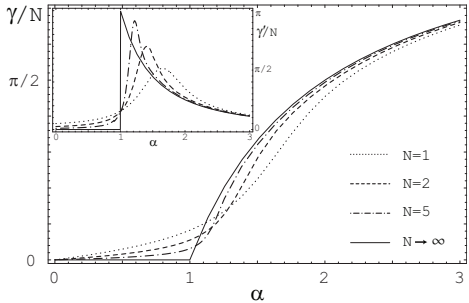}
\caption{The geometric phase $\gamma$ $(\equiv \beta_g)$ of the
ground state and its derivative (inset) for the Dicke model with
respect to the parameter $\alpha$ for different values of qubit
$N$ and the parameter $D=10$. The geometric phase increases with
$\alpha$, and there is a cusplike behavior in the thermodynamic
limit at the critical transition point $\alpha=1$. } \label{Fig5}
\end{figure}

The scaled geometric phase $\beta_g/N$ and its derivative with
respect to the parameter $\alpha$ for $D=10$ is shown in
Fig.\ref{Fig5} \cite{Plastina}. It is evident that the geometric
phase increases with increasing the coupling constant at the
finite qubit number $N$, while in the thermodynamic limit the
geometric phase vanishes when $\alpha <\alpha_{c}$ and has a
cusplike behavior at the critical point $\alpha=\alpha_{c}$. In
addition, the derivative is discontinuous at the critical point.
These results are consistent with the expected behavior of the
geometric phase across the critical point, and therefore we add
another unusual example to the close relation between geometric
phase and quantum phase transition.

\subsection{The Lipkin-Meshkov-Glick model}

Our third example is the Lipkin-Meshkov-Glick (LMG) model
discussed in Ref\cite{Cui}. The LMG was first introduced in
nuclear physics\cite{Lipkin}. The LMG model describes a set of N
qubits coupled to all others with a strength independent of the
position and the nature of the elements and a magnetic field $h$
in the $z$ direction, i.e., the Hamiltonian is given by
\begin{equation}\label{LMG}
H= - \frac{1}{N}(S^2_x + \gamma S^2_y) - h S_z,
\end{equation}
where $\gamma$ is the anisotropy parameter.
$S_{\alpha}=\sum_{i=1}^{N}\sigma^i_{\alpha}/2 (\alpha=x, y, z)$
and the $\sigma_{\alpha}$ is the Pauli operator, N is the total
particle number in this system. The prefactor $1/N$ is essential
to ensure the convergence of the free energy per spin in the
thermodynamic limit. As widely discussed in the literature ( see,
e.g., Ref. \cite{Botet}), this system displays a second-order
quantum phase transition at the critical point $h=1$.

The diagonalization of the LMG Hamiltonian and derivation of the
geometric phase can be obtained by a standard procedure, which can
be summarized in the following steps\cite{Cui}: (i)  perform a
rotation of the spin operators around the $y$ direction, that
makes the $z$ axis along the so-called semiclassical magnetization
\cite{Dusuel} in which the Hamiltonian described in Eq.\ref{LMG}
has the minimal value in the semiclassical approximation. (ii)
Similar to the XY model and the Dicke model, to introduce a
geometric phase of the ground state, we consider a system which
has a rotation $U(\phi)=e^{-i\phi\tilde{S}_z}$ around the new $z$
direction, and then the Hamiltonian becomes $H(\phi)=U^\dagger
(\phi)H U^{\dagger}(\phi)$. (iii) then we use the
Holstein-Primakoff representation,
\begin{eqnarray}\label{hp}
\tilde{S}_z(\phi) &=& N/2 - a^{\dagger}a, \nonumber\\
\tilde{S}^{+}(\phi)&=&(N-a^{\dagger}a)^{1/2}a e^{i\phi},\nonumber\\
\tilde{S}^-(\phi)&=&a^{\dagger} e^{- i\phi}(N-a^{\dagger}a)^{1/2}
\end{eqnarray}
in which $a^{\dagger}$ is bosonic operator. Since the $z$ axis is
along the semiclassical magnetization, $a^{\dagger}a/N\ll 1$ is a
reasonable assumption under low-energy approximation, in which $N$
is large but finite. (iv) the Bogoliubov transformation, which
defines the bosonic operator as $\label{bt} b(\phi)=\cosh x a
e^{i\phi} + \sinh x a^{\dagger}e^{-i\phi}$, where
$tanh2x=2\Gamma/\Delta$ with $\Delta=\sin^2\theta - \frac{\gamma+
\cos^2\theta}{2}+h\cos\theta\nonumber$ and
$\Gamma=\frac{\gamma-\cos^2\theta}{4}.$ These procedures
diagonalize the Hamiltonian to a form
\begin{equation}
H_{diag}(\phi)=Nd+\xi + \Delta^D b^{\dagger}(\phi)b(\phi),
\end{equation}
where $d=- \frac{1}{4}(\sin^2\theta + 2h\cos\theta)$,
$\xi=\frac{\Delta}{2}(\sqrt{1-\epsilon^2}-1)\nonumber$
$\Delta^D=\Delta\sqrt{1 - \epsilon^2}\nonumber$, and
$\epsilon=\tanh2x=2\Gamma/\Delta.$ The ground state
$|g(\phi)\rangle$ is determined by the relation
$b(\phi)|g(\phi)\rangle=0.$ Substituting $b(\phi)$ into the
equation above, one finds the ground state,
\begin{eqnarray}
\nonumber |g(\phi)\rangle &
=&\frac{1}{C}\sum_{n=0}^{[N/2]}\sqrt{\frac{(2n-1)!!}{2n!!}}(-
\frac{e^{-i\phi}\sinh x}{e^{i\phi}\cosh x })^{n-1}\\
 & & \cdot(-\sqrt{2}e^{-i\phi}\sinh x ) |2n\rangle,
\end{eqnarray}
where $n!!=n(n-2)(n-4)\cdots$ and $n!!=1$ for $n\leq0$.
$|n\rangle$ is the Fock state of bosonic operator $a^{\dagger}$
and the normalized constant is
$C^2=\sum_{n=0}^{[N/2]}2\sinh^2x\frac{(2n-1)!!}{2n!!}\tanh^{2(n-1)}x$.

The geometric phase $\beta_g$ of the ground state accumulated by
changing $\phi$ from $0$ to $\pi$ can be derived by the standard
method as shown before, and the final result is give by\cite{Cui}
\begin{equation}\label{g}
\beta_g=\pi\left[ 1 -
\frac{\sum_{n=0}^{[N/2]}2n\frac{(2n-1)!!}{2n!!}\tanh^{2(n-1)}x}
{\sum_{n=0}^{[N/2]}\frac{(2n-1)!!}{2n!!}\tanh^{2(n-1)}x} \right].
\end{equation}

To have some basic ideas about the relation between the geometric
phase and phase transition in the LMG model, the geometric phases
$\beta_g$ as a function of the parameters $(\gamma,h)$ have been
plotted in Fig.6\cite{Cui}. It is notable that the geometric phase
$\beta$, independent of the anisotropy, is divergent in the line
$h=1$, where the LMG model has been proven to exhibit a
second-order phase transition\cite{Botet}. The divergence of
geometric phase itself, rather then the derivative of geometric
phase, shows distinguished character from the XY and Dicke models.
This difference stems from that the collective interaction in the
LMG model, which is absent in the XY model\cite{Cui}.

\begin{figure}
\centering \label{Fig_LMG}
\includegraphics[height=4cm]{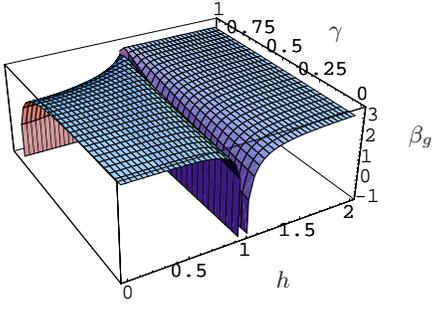}
\put(-40, 5){$h$} \put(-20, 100){$\gamma$} \put(10, 60){$\beta_g$}
\caption{The geometric phase $\beta_g$ of the ground state for the
LMG model as a function of the parameter $(\gamma,h)$ for $N=200$.
The divergence of $\beta_g$ is evident at the critical line
$h_c=1$}
\end{figure}

The scaling behavior of $\beta_g$ has also been studied in Ref
\cite{Cui}. A relatively simply relation $\beta_g\approx- N$ is
obtained there. Furthermore the scaling is independent of
$\gamma$, which means that for different $\gamma$, the phase
transitions belong to the same university class. This phenomenon
is different from the XY model, in which the isotropic and
anisotropic interactions respectively belong  to different
university classes \cite{Zhu2006}.

\section{Pattern II: GP of the test qubit and QPT}

In this section, we consider a test qubit coupled to a quantum
many-body system\cite{Yi,Yuan,Quan}. The Hamiltonian of the whole
system may have the form
\begin{equation}
H=H_t+H_S+H_I,
\end{equation}
where $H_t=\mu \mathbf{B} \cdot \mathbf{\sigma}$ stands for the
Hamiltonian of the test qubit in a general form, $H_S$ represents
the Hamiltonian of a many-body system which we are going to study,
and $H_I$ denotes the coupling between them. We assume that the
quantum system described by $H_S$ undergoes a quantum phase
transition at certain critical points. It is expected that the
geometric phase of the test qubit can be used to identify the
quantum phase transition of the many-body system. A relatively
general formalism to show the close relation between geometric
phase of the test qubit and quantum phase transition of the many
body system has been developed in Ref.\cite{Yi}. For solidness,
here we address a detailed example studied in Ref.\cite{Yuan},
where the many-body system with the quantum phase transition is a
XY spin chain, i.e.,

\begin{equation}
H_t=\mu\sigma^z/2+\nu\sigma^x/2,
\end{equation}

\begin{equation}
 H_S=-\sum_{l=-M}^M \left
(\frac{1+\gamma}{2}\sigma_j^x\sigma_{l+1}^x+\frac{1-\gamma}{2}\sigma_l^y\sigma_{l+1}^y+\lambda\sigma_l^z
\right),
\end{equation}

\begin{equation}
H_I=\frac{\eta}{N}\sum_{l=1}^{N}\sigma^z\sigma_l^z
\end{equation}
where the Pauli matrices $\sigma^{x,y,z}$ and $\sigma_l^{x,y,z}$
denote the test qubit and the XY spin chain subsystems,
respectively. The parameter $\eta$ represents the coupling
strength between the test qubit and all spins (qubits) in the spin
chain. This model is similar to the Hepp-Coleman model\cite{Hepp},
which was initially proposed as a model for quantum measurement,
and its generalization\cite{Nakazato,Sun}.

Following Ref.[15], we assume that the test qubit is initially in
a superposition state
$|\phi_{t}(0)\rangle=c_{g}|g\rangle+c_{e}|e\rangle$, where
$|g\rangle=\left(
\sin\frac{\theta_0}{2},-\cos\frac{\theta_0}{2}\right)
^T$ and $|e\rangle=\left(  \cos\frac{\theta_0}{2},\sin\frac{\theta_0}%
{2}\right)^T$ with $\theta_0=\tan^{-1}(\nu/\mu)$ are ground and
excited states of $H_{t}$, respectively. The coefficients $c_{g}$
and $c_{e}$ satisfy the normalization condition,
$|c_{g}|^{2}+|c_{e}|^{2}=1$. Then the evolution of the $XY$ spin
chain initially prepared in $|\varphi(0)\rangle$, will split into
two branches $|\varphi_{\alpha}(t)\rangle=\exp(-iH_{\alpha
}t)|\varphi(0)\rangle$ ($\alpha=g,e$), and the total wave function
is obtained
as $|\psi(t)\rangle=c_{g}|g\rangle\otimes|\varphi_{g}(t)\rangle+c_{e}%
|e\rangle\otimes|\varphi_{e}(t)\rangle$. Here, the evolutions of
the two branch wave functions $|\varphi_{\alpha}(t)\rangle$ are
driven, respectively, by the two effective Hamiltonians
\begin{equation}
H_{g}=\langle
g|H|g\rangle=H_{S}-\delta\sum_{l=1}^N\sigma_{l}^{z}-\Delta,
\end{equation}

\begin{equation}
H_{e}=\langle e|H|e\rangle=H_{S}+\delta\sum_{l=1}^N
\sigma_{l}^{z}+\Delta,
\end{equation}
where $\Delta=\sqrt{\mu^{2}+\nu^{2}}/2$ and
$\delta=\eta\cos\theta_0/N$. Both $H_{g}$ and $H_{e}$ describe the
$XY$ model in a transverse field, but with a tiny difference in
the field strength. Similar to the method to diagonalize the
standard XY spin chain addressed in the patter I, the ground
states of the Hamiltonians $H_\alpha$  are given by
\begin{equation} |G_\alpha\rangle=\prod_{k=1}^M \left (
\cos\frac{\theta_k^\alpha}{2}|0\rangle_k |0\rangle_{-k} +i
\sin\frac{\theta_k^\alpha}{2}|1\rangle_k|1\rangle_{-k} \right),
\end{equation}
where $\cos\theta_k^\alpha=\epsilon_k^\alpha/\Lambda_k^\alpha$
with
$\Lambda_k^\alpha=\sqrt{\epsilon_{k,\alpha}^{2}+\gamma^{2}\sin^{2}\frac{2\pi
k}{N}}$ and $\epsilon_{k,\alpha}=\lambda-\cos\frac{2\pi
k}{N}+\kappa_{\alpha}\delta$ ($\kappa_g=-\kappa_e=1$).
$|0\rangle_k$ and $|1\rangle_k$ are the vacuum and single
excitation of the $k$th mode, respectively. Here $d_k$ is
similarly defined as the standard XY model (see section 3.1).

Now we turn to study the behaviors of the geometric phase for the
test qubit when the XY spin chain is at its ground state. Due to
the coupling, it is expected that the geometric phase for the test
qubit will be profoundly influenced by the occurrence of quantum
phase transition in spin-chain environment. Since we are
interesting to the quantum phase transition, which is the property
of the ground state, we assume that the $XY$ spin chain is
adiabatically in the ground state $|G_g(\{\theta_{k}^g\})\rangle$
of $H_{g}$. In this case the effective mean-field Hamiltonian for
the test qubit is given by
\begin{eqnarray}
H_{eff}  &=& H_{t}+\langle G_g|H_{I}|G_g\rangle\\
&=& \left(  \frac{\mu}{2}+\frac{2\eta}{N}\sum_{k=1}^{M}\cos\theta_{k}%
^{(g)}\right)  \sigma^{z}+\frac{\nu}{2}\sigma^{x}.
\end{eqnarray}
In order to generate a geometric phase for the test qubit, as
usual, we change the Hamiltonian by means of a unitary
transformation: $U(\phi)=\exp\left(
-i\frac{\phi}{2}\sigma_{z}\right),$ where $\phi$ is a slowly
varying parameter, changing from $0$ to $\pi$. The transformed
Hamiltonian can be written as $H(\phi)
=U^{+}(\phi)H_{eff}U(\phi)$, i.e.,
\begin{equation}
 H (\phi)=\left(  \frac{\mu}{2}+\frac{2\eta}{N}\sum_{k=1}^{M}\cos\theta_{k}%
^{(g)}\right)
\sigma^{z}+\frac{\nu}{2}(\sigma^{x}\cos\phi-\sigma^{y}\sin \phi).
\end{equation}
Then the eigen-energies of the effective Hamiltonian for the test
qubit are given by
\begin{equation}
E_{e,g}=\pm\sqrt{\left(
\frac{\mu}{2}+\frac{2\eta}{N}\sum_{k=1}^{M}\cos
\theta_{k}^{(g)}\right)  ^{2}+\frac{\nu^{2}}{4}}.%
\end{equation}
and the corresponding eigenstates are given by
\begin{equation}
|g\rangle=\left( \begin{array}{l}
\sin\frac{\theta}{2} \\
-\cos\frac{\theta}{2} e^{-i\phi}
\end{array} \right),
|e\rangle=\left(
\begin{array}{l}
\cos\frac{\theta}{2}\\
\sin\frac{\theta}{2}e^{-i\phi}
\end{array}\right),
\end{equation}
where $\sin\theta=\nu/2E_{e}$.

The accumulated  ground-state geometric phase $\beta_g$ for the
test qubit by varying $\phi$ from zero to $\pi$ can be derived
from the standard integral $\int_0^\pi \langle
G_g|\partial_\phi|G_g\rangle d\phi,$ and it is easy to find that
\begin{equation}
\label{T_phase} \beta_{g} =\pi\left( 1+\frac{\mu+4\eta
f(\lambda,\gamma,N)}{\sqrt{[\mu+4\eta f(\lambda
,N)]^{2}+\nu^{2}}}\right),
\end{equation}
where $f(\lambda,\gamma,N)=\frac{1}{N}\sum_{k=1}^{M}\cos
\theta_{k}^{(g)}$. In the thermodynamic limit
$N\rightarrow\infty$, the summation in $f(\lambda,\gamma,N)$ can
be replaced by the integral as follows:
\begin{equation}
\label{F_function}
f(\lambda,\gamma,N)|_{N\rightarrow\infty}=\frac{1}{2\pi}\int_{0}^{\pi}%
\frac{\lambda-\cos\varphi}{\sqrt{(\lambda-\cos\varphi)^{2}+\gamma^{2}\sin
^{2}\varphi}}d\varphi.
\end{equation}

The geometric phase $\beta_g$ and its derivative
$d\beta_g/d\lambda$ with respect to the parameter
$(\lambda,\gamma)$ of the XY model are plotted in Fig.7. As
expected, the nonanalytic behavior of the geometric phase and the
corresponding anomalies in its derivative $d\beta_g/d\lambda$
along the critical lines $\lambda_c=1$ are clear. All these
features are very similar to those in the XY spin chain in patter
I (see section 3.1).

\begin{figure}
\centering \label{fig7}
\includegraphics[height=4cm,width=8cm]{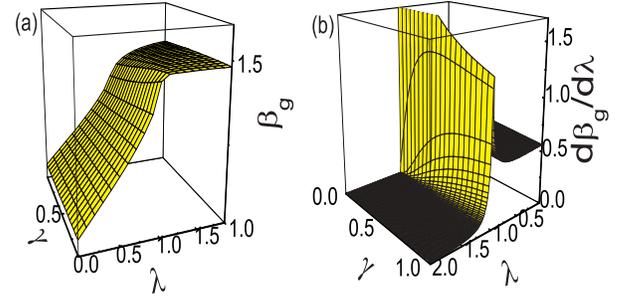}
\caption{(a) Ground-state geometric phase $\beta_g$ of the test
qubit and (b) its derivative $d\beta_g/d\lambda$ as a function of
the spin-chain parameter $(\lambda,\gamma)$. The anomalies for the
derivative of geometric phase is clear along the critical line
$\lambda_c=1$. The other parameters: $\mu=0.1$, $\nu=2$, and
$\eta=0.5$. }
\end{figure}

\begin{figure}
\centering \label{Yuan_2_LMG}
\includegraphics[height=4cm]{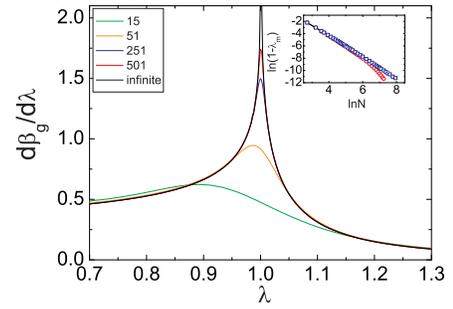}
\caption{The  derivatives $d\beta_g/d\lambda$ for the test qubit
which is coupling to the Ising spin chain ($\gamma=1$), with
respect to the parameter $\lambda$ for different lattice sizes
$N=12,51,251,501,\infty$. With increasing the system sizes, the
maximum becomes more pronounced and the position of the maximum
clearly approaches $\lambda_c=1$ as $N \rightarrow \infty$. The
inset shows the size scaling of the position of the peak occurred
in $d\beta_g/d\lambda$ (circles) and the function
$f(\lambda,\gamma,N)$ (squares). }
\end{figure}

To further understand the relation between GPs and QPTs in this
system, let us consider the case of $XX$ spin model ($\gamma=0$)
in which geometric phase can be analytically derived. In the
thermodynamic limit, the function $f(\lambda,\gamma,N)$ in
Eq.\ref{F_function}  can be derived explicitly for $\gamma=0$ as
$f=1/2-\arccos(\lambda)/\pi$ when $\lambda\leq1$ and $f=1/2$ when
$\lambda>1$. In this case, the geometric phase of the test qubit
is given by
\begin{equation}
\beta_{g}\bigr |_{N\rightarrow\infty}=\left\{
\begin{array}
[c]{l}%
\pi\left(  1+\frac{\mu+2g[1-2\arccos(\lambda)/\pi]}{\sqrt{\left(
\mu+2g[1-2\arccos(\lambda)/\pi]\right)  ^{2}+\nu^{2}}}\right)
\ \ \ \ \ (\lambda\leq1)\\
\pi\left(  1+\frac{\mu+2g}{\sqrt{\left(  \mu+2g\right)
^{2}+\nu^{2}}}\right)
 \ \ \ \ \ \ \ \ \ \ \ \ \ \ \ \ \ \ \ \ \ \ (\lambda>1)%
\end{array}
\right.
\end{equation}
which clearly shows a discontinuity at $\lambda=\lambda_{c}=1$.
The derivative $d\beta_g/d\lambda$ as a function of $\lambda$ for
$\gamma=1$ and different lattice sizes are plotted in Fig. 8
\cite{Yuan}. It is notable that the derivative of geometric phase
is peaked around the critical point $\lambda_c=1$. The amplitude
of the peak is prominently enhanced by increasing the lattice size
of the spin chain. The size dependent of the peak position
$\lambda_m$  for $d\beta_g/d\lambda$ is shown in the inset of Fig.
8. For comparison, the size dependence of the peak position in
$\lambda$ space for the derivative $d f/d\lambda$ are also shown
in the inset (squires). The scaling behavior of
$d\beta_g/d\lambda$ and $d f/d\lambda$ are evident in the figure.
All these features are similar to these exhibit in the XY spin
chain of the patter I. Therefore, we can see that QPTs of the XY
spin chain are faithfully reflected by the behaviors of the
ground-state GP and its derivative of the coupled test qubit.

\section{Summary and concluding remarks}

Quantum phase transition plays a key role in condensed matter
physics, while the concept of geometric phase is fundamental in
quantum mechanics.  However, no relevant relation was recognized
before recent work. In this paper, we present a review of the
connection recently established between these two interesting
fields. Phases and phase transitions are traditionally described
by the Ginzburg-Landau symmetry-breaking theory based on order
parameters and long rang correlation. Recent develops offer other
perspectives to understand quantum phase transitions, such as
topological order, quantum entanglement, geometric phases and
other geometric quantities. Before conclusion, we would like to
briefly address that, rather than geometric phase reviewed in this
paper, the deep relationship between some other geometric
quantities and quantum phase transitions has also been revealed.

Quantum fidelity. Recently an approach to quantum phase
transitions based on the concept of quantum fidelity has been put
forward\cite{Zanardi0,Zhou}. In this approach, quantum phase
transitions are characterized by investigating the properties of
the overlap between two ground states corresponding to two
slightly different set of parameters. The overlap between two
states can be considered as a Hilbert-space distance, and is also
called quantum fidelity from the perspective of quantum
information. A drop of the fidelity with scaling behavior is
observed in the vicinity of quantum phase transition and then
quantitative information about critical exponents can be
extracted\cite{Cozzini,Cozzini1}. The physical intuition behind
this relation is straightforward. Quantum phase transitions mark
the separation between regions of the parameter space which
correspond to ground state having deeply different structural
properties. Since the fidelity is a measure of the state-state
distance, the dramatic change of the structure of the ground state
around the quantum critical point should result in a large
distance between two ground states. The study of QPTs based on
quantum fidelity (overlap) has been reported for several
statistical models\cite{Zanardi0,Zhou,Zanardi_JSM,Gu1,Yi_Hann}. In
addition, the dynamic analogy of quantum overlap is the Loschmidt
echo; it has been shown that the Loschmidt echo also exhibits
scaling behavior in the vicinity of the critical
point\cite{Quan,Quan2,Ou}.

 The Riemannian tensor. It
has been shown that the fidelity approach can be better understood
in terms of a Riemannian metric tensor $g$ defined over the
parameter manifold\cite{Zanardi1}. In this approach, the manifold
of coupling constants parameterizing the system's Hamiltonian can
be equipped with a (pseudo) Riemannian tensor $g$ whose
singularities correspond to the critical regions.

We have presented that one can study quantum phase transitions
from the perspective of some geometric objects, such as geometric
phase, quantum fidelity and the Riemannian tensor. Surprisingly,
All these approaches share the same origin and can be therefore
unified by the concept of quantum geometric tensors. We now
briefly recall the formal setting developed in Ref.\cite{Venuti}.
For each element $\eta$ of the parameter manifold $\mathcal{M}$
there is an associated Hamiltonian $H
(\eta)=\sum_{n=0}^{dim\mathcal{H}} E_n (\eta) |\Psi_n
(\eta)\rangle \langle \Psi_n (\eta)|$ $(E_{n+1}>E_n)$, acting over
a finite-dimensional state space $\mathcal{H}$. If $|\Psi
(\eta)\rangle$ represents the unique ground state of $H(\eta)$,
then  one has the mapping $\Psi_0:
\mathcal{M}\rightarrow\mathcal{H}:\eta\rightarrow |\Psi
(\eta)\rangle$. In this case, one can define a quantum geometric
tensor which is a complex hermitean tensor in the parameter
manifold $\mathcal{M}$ given by \cite{Provost}
\begin{equation}
Q_{\mu\nu} \equiv \langle \partial_\mu \Psi_0|\partial_\nu
\Psi_0\rangle -\langle \partial_\mu \Psi_0|\Psi_0\rangle
\langle\Psi_0|\partial_\nu \Psi_0\rangle,
\end{equation}
where the indices $\mu$ and $\nu$ denote the coordinates of
$\mathcal{M}$. The real part of the quantum geometric tensor $Q$
is the Riemannian metric, while the imaginary part is the
curvature form giving rise to a geometric phase\cite{Venuti}.
Similar to the heuristic argument that we have addressed for the
singularity of Berry curvature in the vicinity of quantum phase
transition, it has been shown that the quantum geometric tensor
also obeys critical scaling
behavior\cite{Venuti,Zanardi0,Zhu2006}. Therefore, viewing quantum
phase transitions from the perspectives of geometric phase and
quantum fidelity can be unified by the concept of quantum
geometric tensor.

In conclusion, we presented a review of criticality of geometric
phase established recently, in which geometric phase associated
with the many-body ground state exhibits universality, or scaling
behavior in the vicinity of the critical point. In addition, we
addressed that one can investigate quantum phase transition from
the views of some typical geometric quantities. The closed
relation recently recognized between quantum phase transitions and
quantum geometric tensor may open attractive avenues and fruitful
dialog between different scientific communities..


\section*{Acknowledgements}

This work was supported by the State Key Program for Basic
Research of China (No. 2006CB921800), the NCET and NSFC (No.
10674049).

\end{document}